\documentclass[trackchanges,twocolumn]{aastex62}

\usepackage{blindtext}

\usepackage{paralist}

\usepackage{amsmath}
\usepackage{color}
\usepackage{mathrsfs}

        {\begin{compactenum}[#1]}{\end{compactenum}}

\newcommand{\uofa}{\affiliation{University of Arizona, Tucson, AZ, USA}}
\newcommand{\ucl}{\affil{The Blackett Laboratory, Imperial College London, London, SW7 2AZ, UK}}
\newcommand{\mullard}{\affil{Mullard Space Science Laboratory, University College London, Holmbury St. Mary, Surrey RH5 6NT, UK}}
\newcommand{\lesia}{\affiliation{LESIA, Observatoire de Paris, Meudon, France.}}

\graphicspath{{./}{figures/}}

\newcommand{\change}[1]{{#1}}

\received{}
\revised{}
\accepted{}

\submitjournal{ApJ (accepted)}

\shorttitle{Linear Stability in the Inner Heliosphere}
\shortauthors{Klein et al.}

\newcommand{\V}[1]{\mathbf{#1}} 

\begin{document}
\title{Linear Stability in the Inner Heliosphere: Helios Re\"evaluated}

\correspondingauthor{K. G. Klein}
\email{kgklein@lpl.arizona.edu}

\author[0000-0001-6038-1923]{Kristopher G. Klein}\uofa
\author[0000-0002-7365-0472]{Mihailo Martinovi\'c}\uofa \lesia
\author[0000-0002-1365-1908]{David Stansby}\mullard
\author[0000-0002-7572-4690]{Timothy S. Horbury}\ucl

\begin{abstract}
  Wave-particle instabilities driven by departures from local
  thermodynamic equilibrium have been conjectured to play a role in
  governing solar wind dynamics.  We calculate the statistical
  variation of linear stability over a large subset of Helios I and II
  fast solar wind observations using a numerical evaluation of the
  Nyquist stability criterion, accounting for multiple sources of free
  energy associated with protons and helium including temperature
  anisotropies and relative drifts. We find that 88\% of the surveyed
  intervals are linearly unstable.  The median growth rate of the
  unstable modes is within an order of magnitude of the turbulent
  transfer rate, fast enough to potentially impact the turbulent
  scale-to-scale energy transfer. This rate does not significantly
  change with radial distance, though the nature of the unstable
  modes, and which ion components are responsible for driving the
  instabilities, does vary.  The effect of ion-ion collisions on
  stability is found to be significant; collisionally young wind is
  much more unstable than collsionally old wind, with very different
  kinds of instabilities present in the two kinds of wind.
  \end{abstract}

\section{Introduction}

Departures of particle velocity distributions from local thermodynamic
equilibrium in the weakly collisional solar wind are frequently
observed (see \cite{Marsch:2012} and \cite{Verscharen:2019} for a
review of such measurements) and are associated with wave-particle
instabilities, a class of interactions that act to move particle
distributions towards equilibrium while simultaneously emitting a
variety of plasma waves. A review of the linear and quasilinear
theory associated with such instabilities can be found in
\cite{Gary:1993} and \cite{Yoon:2017}. The marginal stability
thresholds, surfaces in parameter space beyond which an instability
has a significant growth rate, predicted by such theories have been
shown to constrain observed distributions of parameters in the solar
wind and magnetosphere
\citep{Kasper:2002,Hellinger:2006,Matteini:2007,Bale:2009,Maruca:2012,Verscharen:2013a,Hellinger:2014,Chen:2016,Maruca:2018}.

Though these analytic limits are useful, they typically only focus on
the effect of a single source of free energy, e.g. proton temperature
anisotropy, excluding other sources, e.g. relative drifts between
component distributions or anisotropies of other species. While the
presence of many sources of free energy complicates the application of
analytic limits, the stability of a linearized system can be
determined through the application of the Nyquist instability
criterion\citep{Nyquist:1932}. In this work, we evaluate the stability
of 45,147 solar wind observations made by the Helios spacecraft, a
subset of observations with a bi-Maxwellian fit for the Helium
($\alpha$) component described in \cite{Stansby:2019}, using a numerical
implementation of the Nyquist criterion applied to the hot plasma
dispersion relation for an arbitrary number of relatively drifting
bi-Maxwellian components \citep{Klein:2017}. The Nyquist criterion has
been previously applied to hundreds of solar wind measurements made
by the Wind spacecraft at 1 au \citep{Klein:2018}. The larger number
of observations combined with Helios' coverage of inner heliospheric
radial distances provides an excellent opportunity to characterize how
instabilities behave under different plasma and solar wind conditions.

After reviewing the numerical methodology and the data set, we survey
the calculated linear stability as a function of solar wind and plasma
parameters.  More than 87\% of intervals were found to be linearly
unstable, slightly higher than expectations from simple stability
thresholds.  The median growth rates of the unstable intervals are
considerable, usually around 20-30\% of the turbulent cascade rate.
There is no significant change in the fraction of unstable modes with
increasing distance in the selected data set. The character of the
unstable modes do change as a parcel of plasma moves further from the
Sun, with the $\alpha$ component playing a more significant role in
driving unstable behavior at larger distances.  Highly collisional
intervals are much less unstable, with the $\alpha$ component acting
as the sole source of free energy for the most collisionally processed
intervals.

\section{Data \& Methodology}

\subsection{Methodology}

Instead of using canonical parametric stability thresholds
\citep{Gary:1998,Kasper:2002,Hellinger:2006,Bale:2009}, which
typically depend on only a single source of free energy, we determine
the linear stability for a set of measured parameters by applying a
numerical implementation of the Nyquist instability criterion
(\texttt{plumage}, described in \cite{Klein:2017}). We numerically
evaluate a contour integral over the inverse of the hot plasma
dispersion relation $\det \left[\mathcal{D}(\omega,\V{k};\mathcal{P})
  \right]$,
\begin{equation}
  W_n \left(\V{k},\mathcal{P}\right)= \frac{1}{2 \pi i} \oint \frac{d \omega}
  {\det \left[\mathcal{D}(\omega,\V{k};\mathcal{P}) \right]}.
\end{equation}
The real and imaginary components of the
complex frequency $\omega$ are $\omega_{\textrm{r}}$ and $\gamma$,
$\V{k}$ is the wavevector and $\mathcal{P}$ is the set of
dimensionless plasma parameters describing the system, including
e.g. plasma $\beta$ and $T_\perp/T_\parallel$. As shown by
\cite{Nyquist:1932}, the integer solution of this integral
$W_n(\V{k},\mathcal{P})$ represents the number of linearly unstable
modes supported by the dispersion relation for a particular wavevector
$\V{k}$ at a particular point in parameter space $\mathcal{P}$. As
demonstrated in \cite{Klein:2017}, by iteratively increasing the lower
bound of the integration path from $\gamma=0$ to a finite $\gamma$,
the maximum growth rate $\gamma^{\textrm{max}}(\V{k},\mathcal{P})$ can
be determined as the value of $\gamma$ for which
$W_n(\V{k},\mathcal{P})$ is non-zero, but for any larger value of
$\gamma$, $W_n(\V{k},\mathcal{P})=0$.

This calculation can be performed for any dispersion relation that can
be numerically evaluated; for this work, we focus on the case of a
collection of an arbitrarily number of bi-Maxwellian components
drifting with respect to one another, and use the dispersion relation
numerically evaluated by \texttt{PLUME} \citep{Klein:2015a} Each
component $j$ has a unique temperature parallel $(T_{\parallel,j})$
and perpendicular $(T_{\perp,j})$ to the local magnetic field $\V{B}$,
a density $n_j$, and a bulk velocity $\V{U}_j$, as well as the
intrinsic mass $m_j$ and charge $q_j$ associated with the
component. For this calculation, we use a reference component to
normalize these quantities and produce six dimensionless parameters to
describe each component: $T_{\perp,j}/T_{\parallel,j}$,
$T_{\parallel,\textrm{ref}}/T_{\parallel,j}$, $\Delta
v_{j,\textrm{ref}}/v_{A,\textrm{ref}}$, $n_{j}/n_{\textrm{ref}}$,
$m_{j}/m_{\textrm{ref}}$, and $q_{j}/q_{\textrm{ref}}$.  The Alfv\'en
velocity is calculated using only the reference mass density
$v_{A,\textrm{ref}} = B/\sqrt{4 \pi n_{\textrm{ref}}
  m_{\textrm{ref}}}$ and $\Delta v_{j,\textrm{ref}}$ is the
magnetic-field aligned difference in drift speeds between the
reference species and component $j$.  The global dimensionless
parameters $\beta_{\parallel,\textrm{ref}} = 8 \pi n_{\textrm{ref}}
T_{\parallel,\textrm{ref}}/B^2$ and the reference thermal velocity
normalized by the speed of light $w_{\parallel,\textrm{ref}}/c$
complete our description of the system. Timescales are normalized to
the reference cyclotron frequency
$\Omega_{\textrm{ref}}=q_{\textrm{ref}}B/m_{\textrm{ref}} c$.

\subsection{Data}

While this numerical method has been previously applied to hundreds of
intervals from the Wind spacecraft at 1 au \citep{Klein:2018}, a
statistical assessment of linear stability as a function of solar wind
and plasma parameters requires analyzing a larger set of
data. Additionally, using measurements closer to the Sun than $1$ au
assists the study of radial variations in inferred stability. For
these reasons, we use the recent reanalysis of Helios proton core
\citep{Stansby:2018} and helium \citep{Stansby:2019} measurements,
which produced parallel and perpendicular temperatures, densities, and
drift speeds for the proton core and helium components.  \change{This
  processing of the Helios data did not include a secondary proton, or
  beam, component.} Due to the method used to extract helium
temperature, this data set has a bias toward faster solar wind, and
does not have a uniform coverage of all radial distances; shown in the
top row of Fig.~\ref{fig:stats_1}. \change{All intervals are
  of the same length, that of the integration time for the Helios on
  board electrostatic analyzer, 40.5 seconds.}

The dispersion relation $\det \left[\mathcal{D}(\omega/\Omega_p,\V{k}
  \rho_p;\mathcal{P}) \right]$ for each of the $45,147$ measured
intervals in this data set can be described using
\begin{eqnarray}
  \mathcal{P} =&
\{\beta_{\parallel,p}, \frac{w_{\parallel,p}}{c}, \frac{T_{\perp,p}}{T_{\parallel,p}},
  \frac{T_{\perp,\alpha}}{T_{\parallel,\alpha}}, \frac{T_{\perp,e}}{T_{\parallel,e}},
  \frac{T_{\parallel,p}}{T_{\parallel,\alpha}}, \frac{T_{\parallel,p}}{T_{\parallel,
      e}}, \\
  &
  \frac{n_{\alpha}}{n_{p}}, \frac{n_{e}}{n_{p}}, \frac{\Delta v_{\alpha,p}}{v_{A,p}},
  \frac{\Delta v_{e,p}}{v_{A,p}} \},
  \nonumber
\end{eqnarray}
as well as the intrinsic mass and charge ratios. The subscripts $p$,
$\alpha$, and $e$ indicate proton, Helium, and electron
quantities. Given the observed complexity of electron structure and
the comparatively simple, single bi-Maxwellian model used in this
study, we refrain from any analysis of electron-scale instabilities,
and set $n_e/n_p$ and $\Delta v_{e,p}/v_{A,p}$ to enforce zero net
current and quasi-neutrality. We also assume an isotropic electron
distribution $T_{\perp,e}/T_{\parallel,e}=1$ and set $T_p = (2
T_{\perp,p} + T_{\parallel,p})/3 =T_e$. This leaves us with a seven
dimensional set of varying parameters for our stability model:
\begin{eqnarray}
  \mathcal{P} =&
\left \{\beta_{\parallel,p}, \frac{w_{\parallel,p}}{c}, \frac{T_{\perp,p}}{T_{\parallel,p}},
  \frac{T_{\perp,\alpha}}{T_{\parallel,\alpha}},
  \frac{T_{\parallel,p}}{T_{\parallel,\alpha}},
  \frac{n_{\alpha}}{n_{p}}, \frac{\Delta v_{\alpha,p}}{v_{A,p}} \right \}.
\end{eqnarray}
For each interval with a good $\alpha$ fit in \cite{Stansby:2019},
values for $\mathcal{P}$ are calculated by combining the $\alpha$
properties with bi-Maxwellian proton-core fits from
\cite{Stansby:2018}, supplemented by solar wind parameters from the
publicly available Helios Data Archive \citep{CDAWeb_2018b}.

\subsection{Analysis}

\begin{figure}[h]
  \includegraphics[width=\columnwidth, viewport = 50 25 250 335, clip=true]{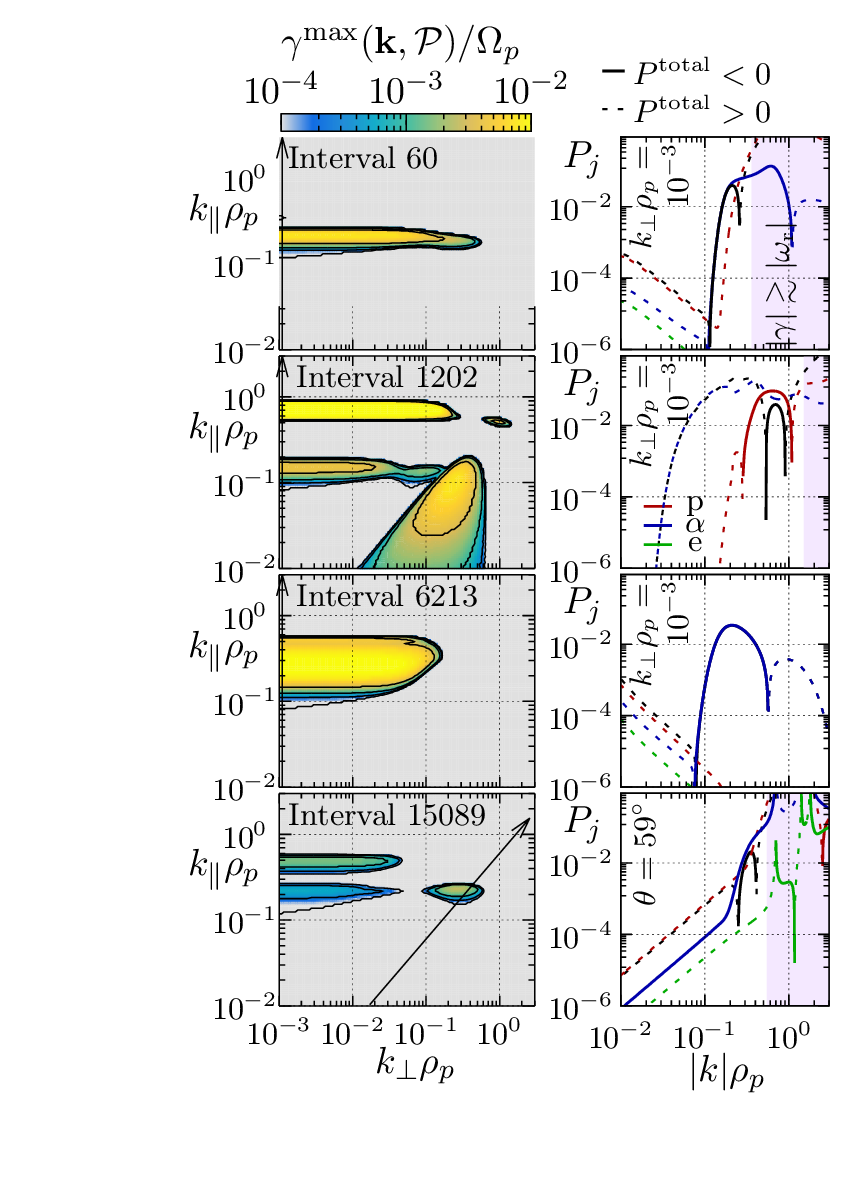}
  \caption{The maximum growth rate
    $\gamma^{\textrm{max}}(\V{k},\mathcal{P})/\Omega_p$ (left column)
    and the power absorption (dashed lines) or emission (solid) for
    the dispersion surface of the fastest growing mode evaluated along
    the arrows in the left column (right). The contributions to the
    total power transfer (black) from the protons (red), $\alpha$s
    (blue), and electrons (green) are separated to identify which
    component drives unstable behavior. Shaded regions indicate
    wavevectors for which the $\gamma \ll \omega_{\textrm{r}}$
    assumption is violated for the selected dispersion surface.}
  \label{fig:example}
\end{figure}

For each interval, we evaluate the growth rate
$\gamma^{\textrm{max}}(\V{k} \rho_p,\mathcal{P})/\Omega_p$ for
wavevectors normalized to the proton gyroradius, $\rho_p = \sqrt{2
  T_{\perp,p}/m_p}$, covering a logarithmically spaced grid, $k_\perp
\rho_p \in [10^{-3},3]$ and $k_\parallel \rho_p \in [10^{-2},3]$,
focusing on instabilities that arise at ion scales. Four examples of
such wavevector grids are shown in the left \change{column} of
Fig.~\ref{fig:example}, with the time of measurement and associated
dimensionless parameters provided in Table~\ref{tab:params}.

\begin{table*}
  \begin{tabular}{|c|c|ccccccc|}
    \hline
    Interval \# & Date & $\beta_{\parallel,p}$ & $w_{\parallel,p}/c$ &  $T_{\perp,p}/T_{\parallel,p}$ & $T_{\parallel,p}/T_{\parallel,\alpha}$ & $T_{\perp,\alpha}/T_{\parallel,\alpha}$ & $n_\alpha/n_p$ & $\Delta v_{\alpha,p}/v_{A,p}$ \\
    \hline
    60 & 1975-02-08 00:41:28 &    $0.642$  &  $8.45\times 10^{-5}$  &  $0.547$ &
    $0.690$ &    $3.38$ &    $0.0315$ &  $-0.0658$ \\
    1202 & 1975-02-09 17:10:25 & $0.240$  &  $1.18\times 10^{-4}$ &  $3.10$ &
      $0.0353$ & $1.49$  &  $0.201$ &  $-0.0136$ \\
      6213 & 1975-02-15 19:08:20 & $0.239$  &  $1.17\times 10^{-4}$ &  $2.26$ &
    $0.0329$ & $0.462$  &  $0.0784$ &  $0.673$ \\
    15089 & 1976-03-19 14:17:10 & $1.86$  &  $2.65\times 10^{-4}$ & $0.606$ &
    $0.252$ &    $1.00$  &   $0.0524$ &  $-1.06$\\
    \hline
  \end{tabular}\\
    \begin{tabular}{|c|ccccccc|}
    \hline
    Interval \# & $\omega^{\textrm{max}}_r/\Omega_p$ & $\gamma^{\textrm{max}}/\Omega_p$ & $k_\perp^{\textrm{max}} \rho_p$ & $k_\parallel^{\textrm{max}} \rho_p$ & $P_p$ & $P_\alpha $ & Polarization\\
    \hline
    60 & $ 0.210$  & $  7.10\times 10^{-3}$ & $ 1.00\times 10^{-3}$ & $ 0.193$ & $ 1.81\times 10^{-3}$ & $ -3.54\times 10^{-2}$ & $-1.00$\\
    1202 & $ -0.606$ & $ 2.33\times 10^{-2}$ & $ 1.00\times 10^{-3}$ & $ 0.684$ & $ -9.23\times 10^{-2}$  & $ 5.60\times 10^{-2}$ & $-1.00$\\
    6213 & $ 0.421$ & $ 1.22\times 10^{-2}$ & $ 1.00\times 10^{-3}$ & $ 0.262$ & $ 1.08\times 10^{-7}$  & $ -2.92\times 10^{-2}$ & $1.00$\\
    15089 & $ -0.142$ & $ 2.81\times 10^{-3}$ & $ 2.80\times 10^{-1}$ & $ 0.238$  & $ 2.25\times 10^{-2}$ & $  -4.26\times 10^{-2}$ & $-0.14$ \\
    \hline
  \end{tabular}
    \caption{Key dimensionless parameters associated with four select
      intervals illustrated in Fig.~\ref{fig:example} and the derived
      wave properties.}
  \label{tab:params}
\end{table*}

For each interval, we determine the largest growth rate within the
specified wavevector range, denoted
$\gamma^{\textrm{max}}(\mathcal{P})/\Omega_p$, and the associated
wavevector $\V{k}^{\textrm{max}} \rho_p$.  While there are frequently
several different types of instabilities associated with a particular
interval, see rows two and four of Fig.~\ref{fig:example}, the fastest
growing mode will dominate the system's dynamics. Using
$\gamma^{\textrm{max}}(\mathcal{P})/\Omega_p$ and
$\V{k}^{\textrm{max}} \rho_p$, we calculate the real frequency of the
fastest growing mode $\omega_{\textrm{r}}(\V{k}^{\textrm{max}}
\rho_p,\mathcal{P})/\Omega_p$ by searching for zeros in $\det
\left[\mathcal{D} (\omega/\Omega_p,\V{k}\rho_p,\mathcal{P}) \right]$
near $\gamma^{\textrm{max}}(\mathcal{P})/\Omega_p$. With
$\omega_r^{\textrm{max}}/\Omega_p$ determined, we calculate the linear
eigenfunctions for electromagnetic field and plasma fluctuations
associated with this normal mode, derived quantities including
magnetic field polarization
\citep{Gary:1993,Krauss-Varban:1994,Schwartz:1996,Klein:Thesis:2013},
and the power absorption or emission per wave-period due to each the
$j^{\textrm{th}}$ plasma component
\citep{Stix:1992,Quataert:1998,Klein:2017b},
\begin{equation}
  P_j = \frac{\gamma_j}{\omega_r} = \frac{\V{E}^* \cdot
    \underline{\underline{\chi}}_j^a \cdot \V{E}}{4 W_{\textrm{EM}}},
\end{equation}
where $\V{E}$ and $\V{E}^*$ are the perturbed electric field and its
complex conjugate, $\underline{\underline{\chi}}_s^a$ the
anti-Hermitian part of the linear susceptibility tensor for component
$j$ evaluated at the normal-mode frequency $\omega_r$, and
$W_{\textrm{EM}}$ is the electromagnetic wave energy. The normal
mode's growth rate is $\gamma/\omega_{\textrm{r}} = \sum_j P_j$;
$P_j<0 (>0)$ denotes energy transfer from plasma component $j$ (waves)
to the waves (plasma component $j$). The wavevector region where the
underlying assumptions for the linear dispersion relation calculation
(i.e. $\gamma < \omega_{\textrm{r}}$) breaks down is shaded in purple
in the right column of Fig.~\ref{fig:example}.

The first row of Fig.~\ref{fig:example} illustrates a typical parallel
resonant instability, with $k_\perp \ll k_\parallel$ and
$k_\parallel \rho_i \lesssim 1$. For this case, the wave resonantly
interacts with the $\alpha$ component; as seen in the sign of
$P_\alpha$ in right column, the power emitted by the $\alpha$s is, for
a narrow range of wavevectors, greater
than the power absorbed by the protons, producing a net transfer of
energy from the charged particles to the electromagnetic
field.\footnote{In all intervals considered, the electrons contributed
  negligibly to the overall energy transfer.}  The fastest growing
mode propagates along $\V{B}$, and the magnetic field fluctuation
polarization in the wave frame at the peak growth rate is left-handed,
indicating the unstable mode is an ion-cyclotron rather than
fast-mode-like wave.

In the second row of Fig.~\ref{fig:example}, we illustrate an example
where protons emit rather than absorb power. Here, the power
absorption by the $\alpha$ component is significant for the
wavevectors where $P_p<0$, limiting the
total growth rate of the unstable mode compared to a proton and
electron description. This mode propagates anti-parallel to
$\V{B}$, but in the wave frame the magnetic field polarization is
left-handed, again representing a ion-cyclotron type wave.

The interval shown in the third row of Fig.~\ref{fig:example}
represents a right-handed, parallel-propagating instability, driven by
the relative drift between ion components. Neither the protons nor
electrons gain energy from or lose energy to the wave, while the
$\alpha$ component emits power over a relatively broad region of
wavevectors compared to the left-handed waves in rows 1 and 2. In a
minority of unstable cases, illustrated the bottom row, an oblique
mode with $k_\perp \gtrsim k_\parallel$ is the fastest growing
mode. Such modes, which represent some version of either the oblique
firehose or mirror instability, have magnetic field polarizations near
zero and lower values for $\omega_{\textrm{r}}$ compared to the
parallel instabilities.

\section{Statistical Analysis of Helios Intervals}

From $45,147$ intervals, we find that $39,695 \ (87.9\%)$ are linearly
unstable with $\gamma^{\textrm{max}}(\mathcal{P})/\Omega_p \geq
10^{-4}$. This decreases to $37,760 \ (83.6 \%)$ and $30,963\ (68.6
\%)$ for thresholds of $\gamma^{\textrm{max}}(\mathcal{P})/\Omega_p
\geq 10^{-3}$ and $10^{-2}$.  Evaluating the stability of the same
intervals using parametric thresholds derived using only proton
temperature anisotropy (e.g. \cite{Verscharen:2016}), and thus
neglecting the effects of the $\alpha$ temperature anisotropy and the
relative drift between the ion components, we find $82.1\%$,
$79.4\%$, and $69.6\%$ of the intervals surpass the
$\gamma/\Omega_p=10^{-4}, 10^{-3},$ and $10^{-2}$ thresholds.  We
hypothesize that this slight decrease in the number of very unstable
intervals may be due to additional power absorption by
the $\alpha$s, as shown in row 2 of Fig.~\ref{fig:example}.

We determine which of the two resolved ion components drives the
fastest growing mode using $P_j$. In most cases, either $P_p< 0$ and
$P_\alpha \gtrsim 0$ ($30,535$ intervals) or $P_\alpha < 0$ and $P_p \gtrsim 0$
($7,845$ intervals), leaving no ambiguity about which component is
emitting power (the component with $P_j < 0$), and which component is
absorbing power $(P_j > 0)$. There are a small number of cases
($1,315$ intervals) where $P_p$ and $P_\alpha$ are less than zero and
are within an order of magnitude of each other; for these cases, we
define the instability as being driven jointly by both ion components.

\begin{figure}[h]
  \hspace*{-1.25cm}
  \includegraphics[width=10cm]{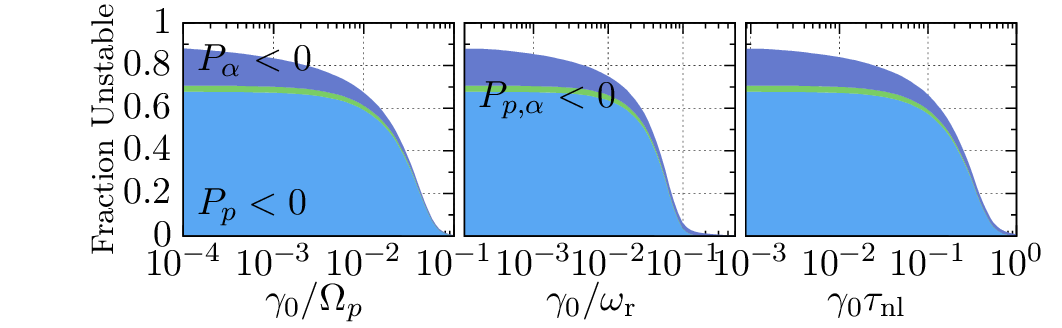}
  \caption{The fraction of intervals that have
    $\gamma^{\textrm{max}}(\mathcal{P}) T$ greater $\gamma_0 T$, for
    $T=\Omega_p^{-1}$ (left), $\omega_{\textrm{r}}^{-1}$ (center), and
    $\tau_{\textrm{nl}}$ (right). Shading indicates whether protons
    (blue) or $\alpha$s (purple) are the primary power emitter, or
    both ion components contribute (green).}
  \label{fig:gamma}
\end{figure}

To further quantify the distribution of growth rates, we plot in
Fig.~\ref{fig:gamma} the fraction of intervals that have
$\gamma^{\textrm{max}}(\mathcal{P})/\Omega_p$ greater than a specified
value, $\gamma_0/\Omega_p$. From this variant of a cumulative
distribution function, we see that most of the unstable modes with
$P_p<0$ have $\gamma^{\textrm{max}}(\mathcal{P})/\Omega_p \approx
10^{-2}$, while most of the unstable modes with $P_\alpha<0$ are more
slowly growing. Overall, $(66,50,33,10) \%$ of the $45,147$ intervals
satisfy $\gamma^{\textrm{max}}(\mathcal{P})
>(0.011, \ 0.022, \ 0.033, \ 0.055)\Omega_p$.

We rescale the growth rates using other characteristic time
scales. Renormalizing the growth rates to $\omega_{\textrm{r}}$,
characterizing how quickly the instability can grow compared to the
wave's propagation, yields a similar distribution to that normalized
to $\Omega_p$, with $(66,50,33,10) \%$ of the total intervals
satisfying $\gamma^{\textrm{max}}(\mathcal{P}) >(0.021, \ 0.038,
\ 0.057, \ 0.088) \omega_{\textrm{r}}$ We note 6 \% of the total
intervals have
$\gamma^{\textrm{max}}(\mathcal{P})/\omega_{\textrm{r}}>0.1$; such
intervals have reached to edge of validity of linear theory.

Renormalizing $\gamma^{\textrm{max}}(\mathcal{P})$ to an estimate for
the critically balanced nonlinear cascade rate $\tau_{\textrm{nl}}\sim
\tau_{\textrm{lin}} \sim \omega_{\textrm{Alfv\'en}}^{-1}$
\citep{Goldreich:1995,Mallet:2015} at scale
$k=|\V{k}^{\textrm{max}}|$,
\begin{equation}
  \tau_{\textrm{nl}} = \frac{\rho_p}{v_A} (k_0
  \rho_p)^{-1/3}(|\V{k}^{\textrm{max}}| \rho_p)^{-2/3}
  \label{eqn:tnl}
\end{equation}
characterizes how quickly the instability can grow compared to the
nonlinear turbulent transfer of energy. To estimate the outer scale
$k_0$, we subdivide Helios magnetometer measurements into twelve hour
intervals, perform a Fourier transform to construct turbulent power
spectra, and identify the spectral break between the energy containing
and inertial ranges.  We average the power spectra into bins as a
function of radial distance and solar wind speed.  Inside each
distance-and-speed bin, the spectral break frequency
$f_{\textrm{break}}$ is found by a linear fitting of the averaged
spectrum on a logarithmic scale using two free parameters,
$f_{\textrm{break}}$ and amplitude of the spectrum at $f=10^{-3}$ Hz,
assuming the spectrum has slopes of $f^{-1}$ and $f^{-5/3}$ above and
below $f_{\textrm{break}}$.  Using these fitted values of
$f_\textrm{break}(R,v_{\textrm{sw}})$, an outer scale for each
interval is calculated, $k_0 = 2 \pi
f_{\textrm{break}}/v_{\textrm{sw}}$, which combined with
$|\V{k}^{\textrm{max}}|$ in Eqn.~\ref{eqn:tnl} yields
$\tau_{\textrm{nl}}$. For all intervals, $(66,50,33,10) \%$ have
$\gamma^{\textrm{max}}(\mathcal{P})\tau_{\textrm{nl}}>(0.102, \ 0.197,
\ 0.297, \ 0.496)$, indicating that for many intervals, the
growth rate is within a factor of a few of the turbulent cascade rate,
and therefore may grow quickly enough to impact the
turbulent scale-to-scale transfer of energy.

As was found at 1 au, parallel instabilities are much more common than
oblique instabilities. We define in this work a parallel instability
as an interval where the fastest growing mode has $k_\perp = 10^{-3}$,
the smallest perpendicular wavevector resolved on our wavevector grid;
oblique instabilities are those with $k_\perp > 10^{-3}$; 39,602
intervals have parallel fastest growing modes, while 93 intervals have
oblique fastest growing modes.  Note that this definition of parallel
and oblique varies from that presented in \cite{Klein:2018}, where the
authors used the wavevector distribution of instabilities combined
with the excess parallel or perpendicular pressure to classify
intervals as kinetic, firehose, or mirror unstable.  We identify many
more than 94 intervals with oblique instabilities, e.g. interval 1202
in Fig.~\ref{fig:example}, but in most cases there exists a faster
growing parallel instability. As the fastest growing mode dominates
the plasma's evolution, we find the definition employed here more
physically meaningful than that used in \cite{Klein:2018}; of the 309
intervals studied in that work, 166 of which where unstable, only 8
have an oblique fastest growing mode.

\begin{figure*}[t]
  \includegraphics[width=18cm]{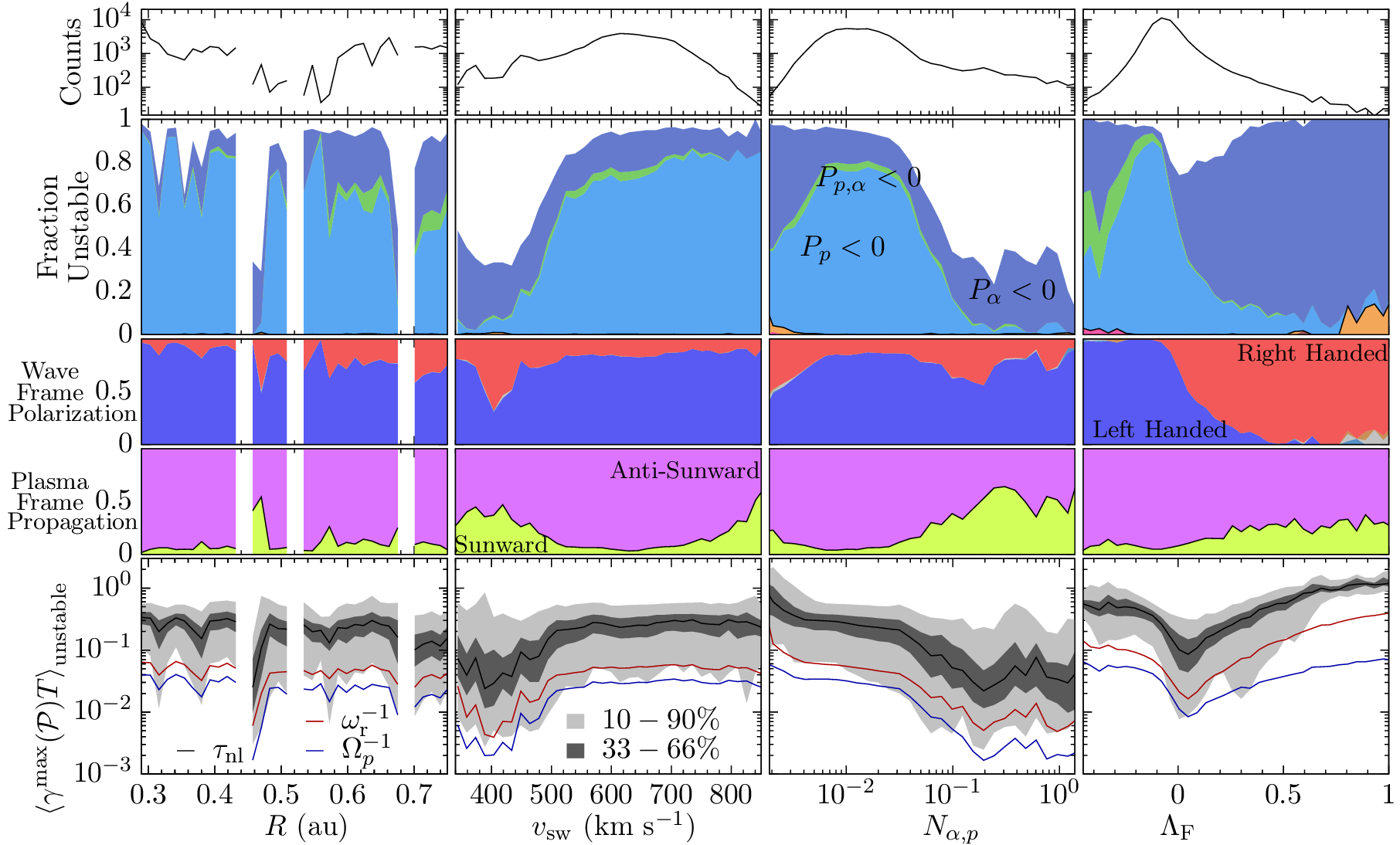}
  \caption{Statistical distributions characterizing the linear
    stability of the 45,147 intervals as a function of radial distance
    $R$, solar wind speed $v_{\textrm{sw}}$, Coulomb number
    $N_{\alpha,p}$, and excess parallel pressure
    $\Lambda_{\textrm{F}}$. The layout is described in the text.}
  \label{fig:stats_1}
\end{figure*}

\begin{figure*}[t]
  \includegraphics[width=18cm]{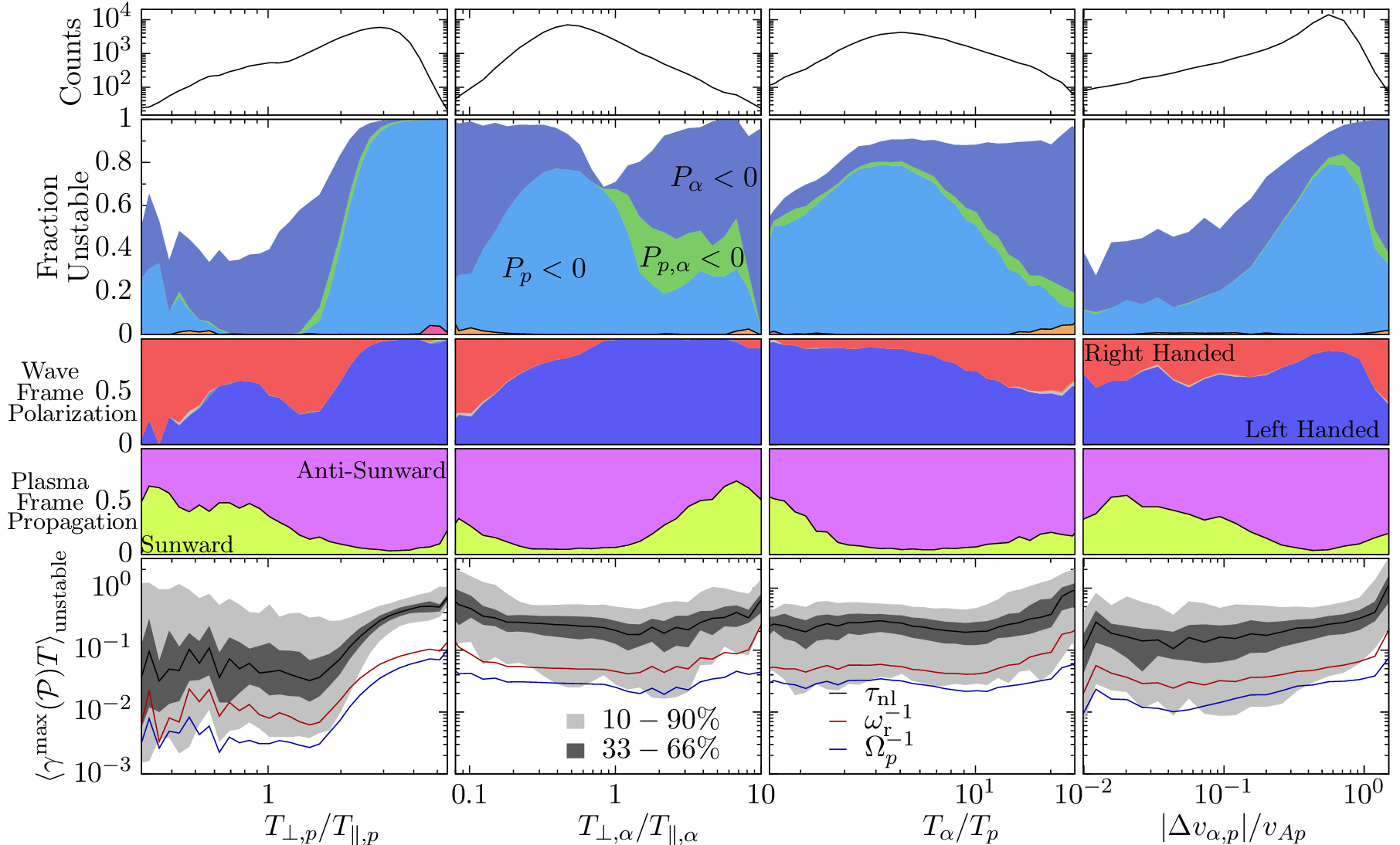}
  \caption{Statistical distributions characterizing the linear
    stability of the 45,147 intervals as a function of temperature
    anisotropies $T_{\perp,p}/T_{\parallel,p}$ and
    $T_{\perp,\alpha}/T_{\parallel,\alpha}$, scalar temperature
    disequilibrium $T_\alpha/T_p$, and normalized drift speed $\Delta
    v_{\alpha,p}/v_{A,p}$. The layout is described in the text.}
  \label{fig:stats_2}
\end{figure*}

In Fig~\ref{fig:stats_1} we subdivide the data set as a function of
radial distance from the Sun $R$, solar wind speed $v_{\textrm{sw}}$,
Coulomb Number $N_{\alpha,p}=\nu_{\alpha,p} R/v_{\textrm{sw}}$
\citep{Feldman:1974,Neugebauer:1976} and excess parallel pressure
\citep{Chen:2016,Kunz:2015}
\begin{equation}
  \Lambda_F = \sum_j \frac{\beta_{\parallel,j}-\beta_{\perp,j}}{2}+
    \frac{\sum_j n_jm_j|\Delta u_j|^2}{(\sum_j n_jm_j)v_A^2}.
\end{equation}
For the calculation of $N_{\alpha,p}$ we use the proton-$\alpha$
collision frequency $\nu_{\alpha,p}$ described in
\cite{Hernandez:1987}. For the calculation of $\Lambda_{\textrm{F}}$,
$\Delta u_j$ is the difference between the bulk velocity of component
$j$ and the center of mass velocity. In Fig~\ref{fig:stats_2}, we
analyze the data as a function of proton temperature anisotropy
$T_{\perp,p}/T_{\parallel,p}$, $\alpha$ temperature anisotropy
$T_{\perp,\alpha}/T_{\parallel,\alpha}$, scalar temperature
disequilibrium $T_\alpha/T_p$, and normalized drift speed $\Delta
v_{\alpha,p}/v_{A,p}$. The rows of Figs~\ref{fig:stats_1} and
\ref{fig:stats_2} are organized as follows: \textbf{top row} the
number of intervals associated with the abscissa co\"ordinate,
\textbf{second row} the fraction of intervals found to be linearly
unstable segregated by angle of $\V{k}^{\textrm{max}}\rho_p$ and $P_j$
(purple is parallel and $P_\alpha<0$, blue parallel and $P_p<0$, green
parallel and both $P_p<0 \ \& \ P_\alpha<0$, orange oblique and
$P_\alpha<0$, and red oblique and $P_p<0$), \textbf{third row} the
distribution of wave-frame magnetic field polarizations (red is
right-handed, blue left-handed, grey nearly zero polarization), \textbf{fourth row} the distribution
of propagation direction in the plasma frame (pink is anti-Sunward,
yellow is Sunward), \textbf{last row} the median value of
$\gamma^{\textrm{max}}(\mathcal{P})/\Omega_p$ (blue),
$\gamma^{\textrm{max}}(\mathcal{P})/\omega_{\textrm{r}}$ (red), and
$\gamma^{\textrm{max}}(\mathcal{P})\tau_{\textrm{nl}}$ (black) drawn from the
subset of unstable intervals; the light and dark grey shading
indicates the ranges which enclose $(10,90) \%$ and $(33,66) \%$ of
the values of $\gamma^{\textrm{max}}(\mathcal{P})\tau_{\textrm{nl}}$.

The fraction of linearly unstable intervals is relatively constant as
a function of radial distance from the Sun.\footnote{The three white
  stripes in the left column of Fig.~\ref{fig:stats_1} represent
  distances with too few measurements to enable a statistical study.}
With increasing distance, the fraction of intervals where the $\alpha$
\change{emission of power drives the fastest growing mode} increases
from a few percent near 0.3 au to more than 20 \% at 0.7 au. A similar
transition from uniformly left-handed waves to a minority of
right-handed waves with increasing distance is seen. The interval
between $0.46$ and $0.48$ au is an unusual subset of the data,
entirely composed of slow wind as opposed to all other distances,
which are dominated by fast wind intervals.  The median growth rate,
normalized to $\Omega_p^{-1}$, $\omega_{\textrm{r}}^{-1}$, or
$\tau_{\textrm{nl}}$, is relatively constant with distance, decreasing
by less than a factor of two over those radial distances.

More significant variations arise with solar wind speed; 90\% of fast
wind intervals with $v_{\textrm{sw}}>550$ km s$^{-1}$ are unstable
with a median growth rate of $\left<\gamma^{\textrm{max}}(\mathcal{P})
\tau_{\textrm{nl}} \right> \approx 0.3$ and are predominately
left-handed. Slower wind is much less unstable, with slower growth
rates $\left<\gamma^{\textrm{max}}(\mathcal{P}) \tau_{\textrm{nl}}
\right> \approx 0.03$. A larger fraction of the unstable modes
in the slow wind propagate toward the Sun. There is a noticeable uptick
in Sunward propagation for wind with $v_{\textrm{sw}}>700$ km
s$^{-1}$, though there are relatively few measurements satisfying this
criterion. For slower wind, more of the unstable intervals are
driven by $\alpha$s than protons.

Similar variations are seen with Coulomb Number. The collisionally
oldest wind, intervals with $N_{\alpha,p}\gtrsim 1$, has fewer
unstable intervals and these instabilities are driven by $\alpha$s
rather than protons and are evenly divided between Sunward and
anti-Sunward propagation. In collisionally younger wind, a majority of
intervals are unstable with protons serving as the main instability
driver.  In the collisionally youngest wind, intervals with
$N_{\alpha,p}\lesssim 5 \times 10^{-3}$, the $\alpha$s drive
approximately half of the unstable modes and the median growth rate is
enhanced. This variation, not observed in the fastest solar, may be a
signature of Helium driven far out of thermal equilibrium in the
near-Sun region of preferential minor ion heating hypothesized by
\cite{Kasper:2017} and \cite{Kasper:2019}, a conjecture that will be
investigated with observations closer to the Sun.

Variations with $\Lambda_{\textrm{F}}$ are stark. For intervals with
$\Lambda_{\textrm{F}}>0$, representing intervals with excess parallel
pressure, the instabilities are generally right-handed and driven by
the $\alpha$ component. This is expected, as $T_{\perp p}/T_{\parallel
  p}$ is generally greater than one, the excess parallel pressure is
driven by the drifting $\alpha$ component. For $\Lambda_{\textrm{F}}
\approx 1$, nearly 5 \% of the intervals have an oblique fastest
growing modes. For $\Lambda_{\textrm{F}}<0$, representing cases with
greater perpendicular than parallel pressure, the instabilities are
left-handed, with contributions from protons, $\alpha$s, or both
ion components, to the unstable growth. For significant deviations from
$\Lambda_{\textrm{F}}=0$ the median growth rates are enhanced.

Unsurprisingly, intervals with large proton temperature anisotropies
are associated with instabilities primarily driven by protons; the modes
are left-handed and anti-Sunward propagating.  Fewer intervals with
$T_{\perp,p}/T_{\parallel,p}<2$ are unstable, and of those that are,
a larger fraction of both right-handed magnetic fluctuations
and (plasma-frame) Sunward propagation orientations. These intervals
also have generally slower growth rates compared to large
anisotropy intervals.

The median growth rates are relatively constant with respect to
$T_{\perp,\alpha}/T_{\parallel,\alpha}$. Departure from
$T_{\perp,\alpha} = T_{\parallel,\alpha}$ leads to an increase in the
fraction of unstable modes. Intervals with $T_{\perp,\alpha} >
T_{\parallel,\alpha}$ are dominantly left-handed, with an admixture of
Sunward and anti-Sunward waves; a significant fraction of these
intervals have both the protons and $\alpha$s emitting
power. Intervals with $T_{\perp,\alpha} < T_{\parallel,\alpha}$ can be
right or left-handed.

\change{Intervals with $T_\alpha \gg T_p$} have the $\alpha$ component
as the primary power emitter. As the \change{ion-temperature ratio
  decreases from $T_\alpha \gg T_p$ to $T_\alpha \gtrsim 4 T_p$}, the
protons drive a larger fraction of the instabilities, \change{though}
$90 \%$ of the intervals remain unstable. Below $T_\alpha/T_p \approx
4$ the fraction of unstable intervals decreases. The median growth
rate is fairly constant \change{regardless of $T_\alpha/T_p$}, with a
slight uptick for $T_\alpha/T_p>10$, though intervals with such
extremely large disequilibrium are relatively rare.

Intervals with relatively weak drift speeds, $\Delta
v_{\alpha,p}/v_{A,p} \lesssim 0.1$ are less unstable than intervals
with faster drift speeds, both in terms of fraction of intervals
unstable and the median growth rate. For the slower drifting cases,
both $\alpha$s and protons contribute to the instabilities, and
approximately equal admixtures of right and left-handed and Sunward
and anti-Sunward waves can be found. As the normalized drift speed
grows toward unity, the unstable waves become preferentially
\change{left-handed} and anti-Sunward propagating, with the protons as the
main driver of unstable growth. This changes for $\Delta
v_{\alpha,p}/v_{A,p} \gtrsim 1.0$, where more of the unstable waves are
right-handed, the $\alpha$s significantly
contribute to unstable growth, and the median growth rate
increases significantly.

\begin{figure*}[ht]
  \includegraphics[width=18cm]{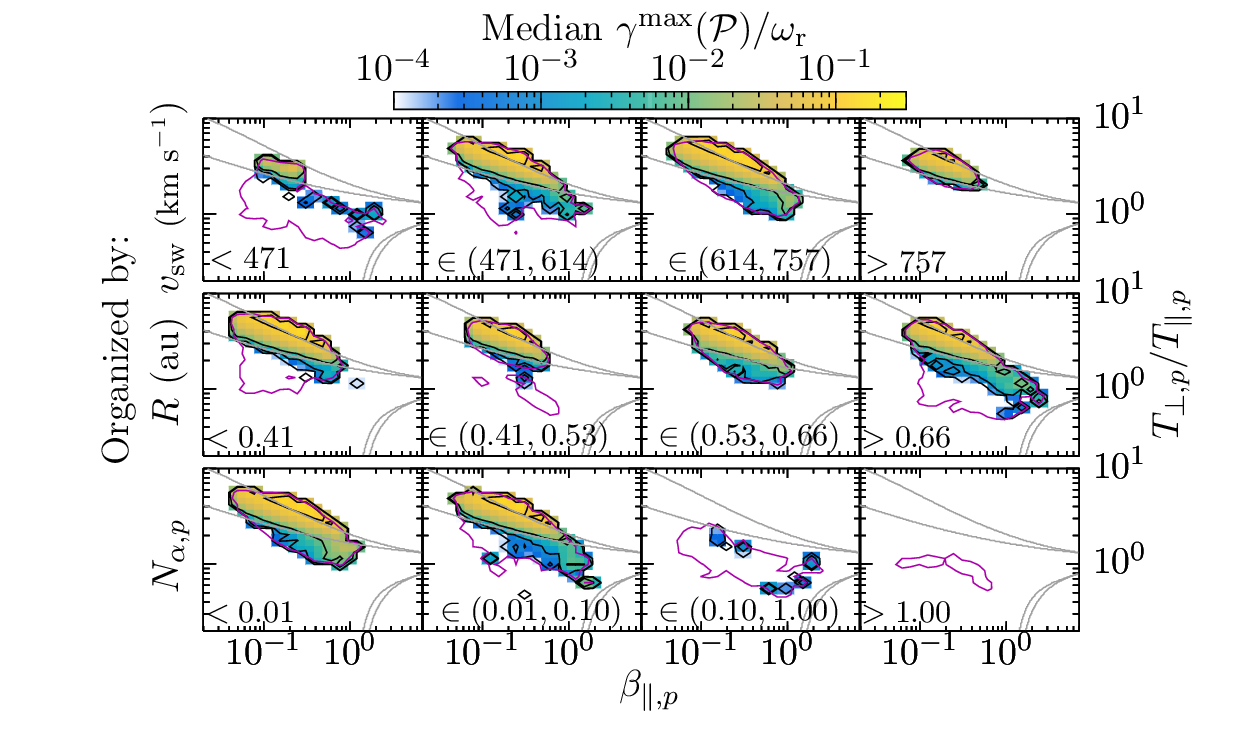}
  \caption{Median Growth rate
    $\gamma^{\textrm{max}}/\omega_{\textrm{r}}$ as a function of
    $\beta_{\parallel,p}$ and $T_{\perp,p}/T_{\parallel,p}$ for
    subsets of the overall data subdivided into ranges of
    $v_{\textrm{sw}}$ (top row), $R$ (center), or $N_{\alpha,p}$ (bottom). The
    pink lines indicate regions for which there are at least 10
    intervals falling into the histogram bin. The grey lines represent
    the proton temperature anisotropy marginal stability thresholds
    for $\gamma/\Omega_p = 10^{-2}$ taken from \cite{Verscharen:2016}.
  }
  \label{fig:quartile}
\end{figure*}

\begin{figure}[h]
  \includegraphics[width=8.75cm]{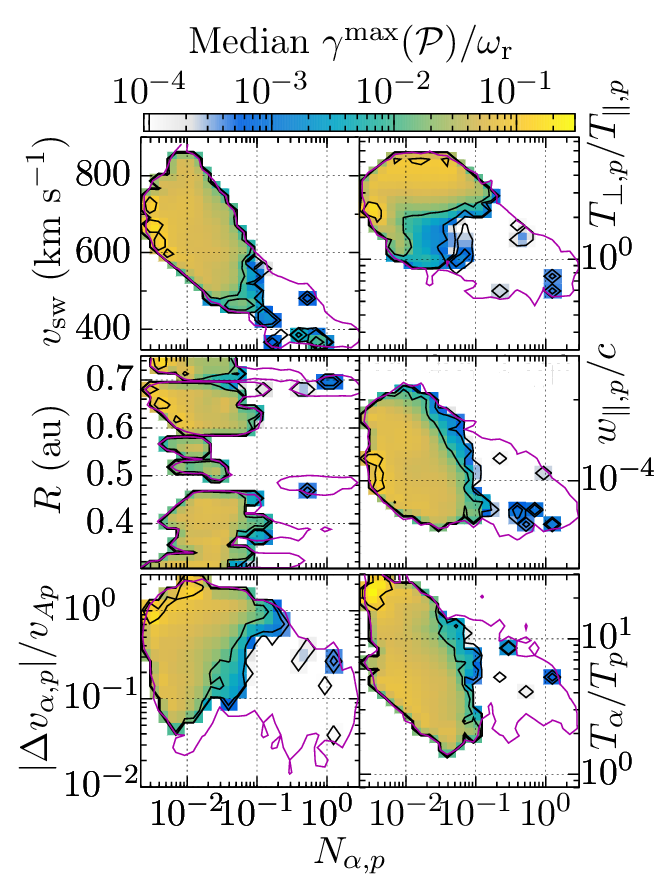}
  \caption{Median Growth rate
    $\gamma^{\textrm{max}}/\omega_{\textrm{r}}(\mathcal{P})$ as a function of
    Coulomb number $N_{\alpha,p}$ and solar wind speed
    $v_{\textrm{sw}}$ (top left), radial distance $R$ (middle left),
    ion drift speed (bottom left), proton temperature anisotropy
    $T_{\perp,p}/T_{\parallel,p}$ (top right), normalized proton
    thermal speed $w_{\parallel,p}/c$ (middle right), or ion
    temperature disequilibrium $T_{\alpha}/T_p$ (bottom right).}
  \label{fig:drift}
\end{figure}

We next calculate the median growth rate
$\gamma^{\textrm{max}}/\omega_{\textrm{r}}(\mathcal{P})$ as a function
of two dimensionless parameters, $\beta_{\parallel,p}$ and
$T_{\perp,p}/T_{\parallel,p}$. In Fig.~\ref{fig:quartile}, the data
are further subdivided into groups based on $v_{\textrm{sw}}$, $R$, or
$N_{\alpha,p}$. Unlike in Figs.~\ref{fig:stats_1} and
\ref{fig:stats_2}, the calculation of the median value in
Figs.~\ref{fig:quartile} and \ref{fig:drift} includes both stable and
unstable intervals, allowing the determination of \change{whether} a
typical interval at a particular point in parameter space is stable or
unstable. As found in \cite{Matteini:2007}, faster wind has larger
proton temperature anisotropies and lower $\beta_{\parallel,p}$, while
slow \change{wind} is more widely distributed throughout the
$(\beta_{\parallel,p},T_{\perp,p}/T_{\parallel,p})$ plane. We see a
spreading of the observed parameter distribution from low
$\beta_{\parallel,p}$ and large $T_{\perp,p}/T_{\parallel,p}$ toward
higher $\beta_{\parallel,p}$ and lower $T_{\perp,p}/T_{\parallel,p}$
with increasing distance, though as seen in Fig.~\ref{fig:stats_1} the
fraction of unstable intervals remains fairly constant with
$R$. Observations that do not surpass the proton-temperature-
anisotropy-only marginal stability thresholds, as previously reported
in \cite{Marsch:2004}, \cite{Matteini:2007}, \cite{Stansby:2019} using
Helios data, and as seen at 1
au\citep{Kasper:2002,Hellinger:2006,Bale:2009} \change{are driven by
  $\alpha$ associated sources of free energy.}

The variation with Coulomb number is striking. For collisionally young
intervals, $N_{\alpha,p}<0.1$, typical values for
$T_{\perp,p}/T_{\parallel,p}$ are larger and regardless of
$\beta_{\parallel,p}$ and $T_{\perp,p}/T_{\parallel,p}$, the typical
interval is unstable. For collisionally old intervals,
$N_{\alpha,p}>0.1$, unstable intervals are much less common, with the
median interval being stable.

To better constrain the role of collisional processing in the
evolution of the solar wind, we calculate the median growth rate
$\gamma^{\textrm{max}}/\omega_{\textrm{r}}(\mathcal{P})$ as a function
of $N_{\alpha,p}$ and $v_{\textrm{sw}}$, $R$, $\Delta
v_{\alpha,p}/v_{Ap}$, $T_{\perp,p}/T_{\parallel,p}$,
$w_{\parallel,p}/c$ (a proxy for the proton temperature
$T_{\parallel,p}$) and $T_{\alpha}/T_p$, shown in
Fig.~\ref{fig:drift}.  In all the joint distributions, we see the same
pattern; for collisionally young solar wind with $N_{\alpha,p}\lesssim
0.1$ the median interval is linearly unstable, with relatively robust
growth rates of
$\gamma^{\textrm{max}}/\omega_{\textrm{r}}(\mathcal{P}) \gtrsim 3
\times 10^{-2}$. For collisionally older wind, most intervals are
linearly stable. From these panels, we also clearly see that the
intervals with the hottest $\alpha$s and the largest drift speeds are
associated with the lowest values of $N_{\alpha,p}$.

\section{Conclusions}

The fraction of unstable intervals found in this data set of $45,147$
Helios observations mostly drawn from the fast solar wind, $87.9 \%$,
is larger than that reported from a much smaller set of 309
observations from the Wind spacecraft at 1 au, $53.7\%$. The larger
number of observations in the processed Helios data set enables a
detailed statistical study of the parametric dependence of
linear instabilities in the solar wind. We find that the occurrence
rate of instabilities does not drastically change from $0.3$ to $0.7$
au, but that the nature of the instabilities changes with increasing
distance, with the $\alpha$ component playing a more
significant role in driving unstable modes at larger distances. The
median growth rates $\gamma^{\textrm{max}}(\mathcal{P})$ are
relatively constant with $R$, with $\gamma^{\textrm{max}}(\mathcal{P})
\tau_{\textrm{nl}} \sim 0.3$, indicating that the growth rates of the
instabilities are a non-negligible fraction of the turbulent cascade
rate at $k^{\textrm{max}}\rho_p$, and thus may grow quickly enough to
impact the turbulent cascade.

It is clear that collisional processing plays an important role in
modifying the particle velocity distributions and reducing the occurrence of
linear instabilities, seen in Figs.~\ref{fig:stats_1},
\ref{fig:quartile}, and \ref{fig:drift}. The continued presence of
$\alpha$-driven instabilities for values of $N_{\alpha,p} \gtrsim 0.1$
while there are relatively few proton-driven instabilities may be
indicative of the faster proton-proton collision rate removing sources
of proton free energy more rapidly than $\alpha$-proton collisions.

It is unclear from these results how this dependence on collisions is
compatible with the isotropization that should accompany linear
wave-particle instabilities, seen for instance in nonlinear
simulations \citep{Hellinger:2003,Hellinger:2008}, \change{though}
evidence for combined action of collisions and instabilities has
arisen from quasilinear models (e.g. \cite{Yoon:2019}).  One
conjecture could be that as the instabilities return the velocity
distributions to marginal stability, finite amplitude fluctuations
could reignite unstable behavior, the so-called fluctuating anisotropy
effect \citep{Verscharen:2016}. This oscillation between unstable and
stable states would only be arrested by the much slower process of
collisional isotropization, which brings the equilibrium distribution
closer to a Maxwellian state, requiring larger amplitude fluctuations
to push the plasma unstable. \change{Ongoing turbulent heating leading
  to increased perpendicular temperatures
  (e.g. \cite{Hellinger:2013,Matteini:2013}) may also play a role in
  keeping distributions pinned near marginal stability in the
  expanding solar wind.}

We emphasize that the processing of the \cite{Stansby:2019} data
set intentionally did not include a proton beam component; this lack
of a proton beam may underestimate the excess parallel pressure in the
system, and may significantly impact a given interval's
stability. \cite{Klein:2018} found at 1 au that $68.8 \%$ of intervals
for which a proton core, beam and $\alpha$ component was resolved were
unstable, while only $29.0 \%$ of intervals with only a proton core
and $\alpha$ component were unstable. Extending this instability
analysis to large sets of measurements that characterize the proton
core, proton beam, and $\alpha$ components at 1 au, in the inner
heliosphere (e.g. the recently reprocessed ion dataset from
\cite{Durovcova:2019}), and at much closer radial distances (e.g. with
Parker Solar Probe \citep{Fox:2015}), will be necessary to improve our
understanding of wave-particle interactions in the inner heliosphere.

Extending our analysis to include departures of the velocity
distribution from a bi-Maxwellian model (e.g. \cite{Verscharen:2018})
will be essential to determining if the solar wind actually supports
these instabilities, or if many of them would be suppressed if
realistic phase-space densities are used in the calculation of the
plasma response (e.g. \cite{Isenberg:2012}).

\change{The authors thank the referee for comments on the submitted
  manuscript.} K. Klein is supported by NASA ECIP Grant 80NSSC19K0912.
D. Stansby is supported by STFC grant ST/S000240/1.  T. Horbury is
supported by STFC grant ST/S000364/1.  An allocation of computer time
from the UA Research Computing High Performance Computing at the
University of Arizona is gratefully acknowledged.

\bibliographystyle{apj}

\end{document}